\input harvmac
\input epsf
\def\thetaswap{^{^{\leftrightarrow}}\hskip -9.5pt \theta}
\def\tthetaswap{^{^{\leftrightarrow}}\hskip -7.5pt \theta}

%%%%%%%%%%%%%%%%%%%%%%%%%%%%%%%%%%%%%%%%%%%%%%%
\lref\douglasI{M.R.~Douglas, hep-th/9901146.}
\lref\douglasII{M.R.~Douglas, hep-th/9910170.}
\lref\azm{L.~Alvarez-Gaume, D.Z.~Freedman and S. Mukhi,
Ann. Phys. {\bf 134}, 85 (1981).}
\lref\grisaru{M.T~Grisaru, A.M.E.~Van~de~Ven and D.~Zanon,
Nucl. Phys. {\bf B277} 388, 409 (1986).}
\lref\sennem{D.~Nemeschansky and A.~Sen, Phys. Lett. {\bf 178B} 365
(1986).}
\lref\groswit{D.J.~Gross and E.~Witten, Nucl. Phys. {\bf B277} 1
(1986).}
\lref\west{West, {\it Introduction to Supersymmetry and Supergravity} 
World Scientific (1986).}

\vskip -10pt
\Title{}
{\vbox{
\centerline{The D0-brane metric in $N$ = 2 sigma models}
}} 
\vskip-10pt
\centerline{{\bf Thomas Wynter}\footnote{$^\circ$}{{\tt
wynter@wasa.saclay.cea.fr}}
}
\centerline{{\it SPhT, CEA-Saclay,}}
\centerline{{\it 91191 Gif-sur-Yvette, France}}

\vskip .6in
\centerline{{\bf Abstract}}
\vskip .15in
\baselineskip10pt{
\hskip -19.5pt We investigate the physical 
metric seen by a D0-brane 
probe in the background geometry of an $N=2$ sigma model. The metric is
evaluated by calculating the Zamolodchikov metric for the disc two point
function of the boundary operators corresponding to the 
displacement of the D0-brane boundary. At two loop order we show that 
the D0 metric receives an $R^2$ contribution.}
\bigskip
\hfill SPhT 00/115

%\draft
\Date{June 2000}

\baselineskip=16pt plus 2pt minus 2pt
\bigskip

%%%%%%%%%%%%%%%%%%%%%%%%%%%%%%%%%%%%%%%%%%%%%%%%%%%%%%
\newsec{Introduction} 

D-branes have provided new insights into the meaning of quantum
geometry.  Used as local probes of string/M theory they have led to
the concept of D-geometry, the particular geometry seen by a D-brane
(see \douglasI\douglasII for reviews). In this article we investigate
the D-brane metric, the metric seen by a D0 brane probe in curved
space. We will work in the classical string limit, $g_s=0$, and will
focus on a background Kahler geometry provided by an $N=2$ non-linear
sigma model.  The calculations will be performed as a perturbative
expansion in $l_s/l_R$ where $l_s$ is the string length and $l_R$ is
the typical curvature radius of the background geometry.

The sigma model metric must satisfy certain equations of
motion to provide a conformally invariant theory. They can be written
in terms of a powers series in $l_s^2$ with the powers of $l_s^{2n}$
arising from an n loop calculation. At lowest order the metric must be
Ricci flat. On general grounds it was known that for a Ricci flat
metric the two loop contribution to the beta function must also be
zero since, for a Kahler metric, the only allowed tensors of the
correct order vanish for Ricci flat metrics. In particular terms such
as $R^2$ (which occur for the bosonic theory) cannot be generated
since they cannot arise from a Kahler potential. Specific calculations
showed that this was indeed the case \azm\ and were eventually pushed
out to four loop order \grisaru . The results being that up to three
loops the beta function vanishes for Ricci flat metrics, but at four
loops there is an $R^4$ contribution which is non-zero for Ricci flat
metrics. Ricci flat metrics are thus a first approximation to the
allowed background metrics of string theory.  Starting from a Ricci
flat metric one can perturbatively (in $l_s^2$) construct a finite
metric satisfying the four loop beta function equation \sennem . It is
always possible to find finite globally well defined non-Ricci flat
corrections to the Kahler potential whose one loop divergences cancel
the divergences from higher loops.

There is nevertheless an ambiguity in the definition of the sigma
model metric. At each order in $l_s^2$ counter-terms are added to
cancel the divergences. Nothing in the renormalization procedure,
however, determines the finite part of these counter-terms. They can
in principle be any covariant tensor of the correct order constructed
from the metric. These counterterms might themselves then lead to
divergences, and can thus alter the beta function and hence the
equation of motion that the sigma model metric must satisfy. There is
nevertheless a physical metric seen by the string. Its equations of
motion are determined by string theory scattering amplitudes. The
calculations of \grisaru\groswit\ showed that, at least to four loop
order, the procedure of minimal subtraction led to beta function
equations of motion for the sigma model metric identical to those
deduced from string scattering amplitudes. The finite counter term
procedure of \sennem\ showed how to construct a finite sigma model
metric satisfying the string scattering amplitude equations of
motion. This metric can be called the``physical'' sigma model metric.

The question addressed in this paper is the relation between this
physical sigma model metric and the metric seen by a D0 brane probe.

The low energy effective action for the
motion of a D0-brane in curved space is of the form
\eqn\DzeroLeff{
S=\int dt g^{D}_{ij}\partial_t X^i \partial_t X^j.
}
Since we are looking at the classical string limit, $g_s=0$, we will be
considering the disc amplitude.  The metric $g^D$ seen by a D0-brane
is a well defined physical metric, given in terms of the disc two
point function for the boundary operators ${\cal O}_i = g_{ij}\partial_n
X^j$ corresponding to shifting the D0-brane boundary\douglasI (The derivative
$\partial_n$ is the derivative normal to the boundary).  Specifically
we have
\eqn\Zamet{
<{\cal O}_i(x_1){\cal O}_j(x_2)>=g^D_{ij}{1\over 2\pi(x_1-x_2)^2}
}
The $1/(x_1-x_2)^2$ dependence is determined on dimensional grounds.
The Zamolodchikov metric $g^D_{ij}$ is the metric on the moduli space
for the D0-brane with the moduli space being the position of the D0-brane
in the curved target space. A heuristic way to understand the
connection between the metric $g^D_{ij}$ in \Zamet\ and the metric
appearing in the low energy effective action is that the Zamolodchikov
metric gives the normalization for the states created by the operators
${\cal O}^i$.

For the calculation of the Zamolodchikov metric we will take as our
bulk CFT an $N=2$ sigma model whose metric satisfies the beta function
equations of motion. The sigma model metric will thus be written as a
power series in $l_s^2$. To order $l_s^4$ the metric is Ricci
flat.  Although the calculations resemble those used to evaluate the
sigma model beta function, there are important differences. For the
sigma model beta function calculation it is only the divergent
contributions that are important, whereas for the Zamolodchikov metric
it is the finite terms that are physically relevant (the vanishing of
the divergent terms is assured by the fact that metric satisfies the
beta function equations of motion). It is thus not at all obvious that
the Zamolodchikov metric will be identical to the sigma model metric.
There are, for example, a priori no reasons why the metric should be a
Kahler metric and hence no a priori reason why at two loop order there
cannot be terms in the metric of the form $R^2$ which do not vanish
for Ricci flat metrics. Below the Zamolodchikov metric is calculated
out to two loop order. It is found that there is such a
contribution. This leads to a D0 brane metric that is neither Ricci
flat (at order $l_s^4$) nor derivable from a Kahler potential. It is
thus a physically different metric from that seen by the string.

We start in section 2 with a brief overview of Kahler geometry and the
background field method.  Calculations are greatly simplified by the
use of the superfield formalism (see for example \west ). The fact
that we have a boundary leads to slight modifications in the
propagators and superderivative/propagator identities from the case
without a boundary. Section 3 is devoted to determining these
differences and setting up the Feynman rules. In section 4 we
calculate the Zamolodchikov metric up to two loops. 

\newsec{The $N=2$ action and the background field method}

The sigma model action, written in terms of chiral 
$\Phi^I(z,\theta,\bar{\theta})$ 
and antichiral 
$\bar{\Phi}^{\bar{J}}(z,\theta,\bar{\theta})$ 
superfields, is given by
\eqn\Kahlact{
S=\int\,d^2z\,d^4\theta\,K(\Phi,\bar{\Phi}).
}
where $K(\Phi,\bar{\Phi})$ is the Kahler potential and the bosonic
components of $\Phi$ and $\bar{\Phi}$ are coordinates on the Kahler
manifold. The world sheet topology is that of the disc with boundary
mapped to the real axis and the integral over $d^2z$ defined over the
upper half plane.

Calculations will be performed using the background field method.  We
start with a classical string world sheet. Since the world sheets we
are considering have the topology of a disc they will classically collapse
down to a single space time point. We write the fields as a constant
part $\Phi_{cl}$, corresponding to this space time point plus a
quantum part $\Phi(z,\theta)$~: 
\eqn\bquflds{
\Phi_{total}(z,\theta) = \Phi_{cl} + \Phi(z,\theta).
}
The Kahler potential is then expanded as a power series in the quantum
fields. 
\eqn\Kahlmet{\eqalign{
K(\Phi_{total},\bar{\Phi}_{total})=
&\,\,\,\,K_{I\bar{J}}\Phi^{I}\bar{\Phi}^{\bar{J}}\cr
&\hskip -10pt+{1\over 2!}
K_{IJ\bar{K}}\Phi^{I}\Phi^{J}\bar{\Phi}^{\bar{K}}
+{1\over 2!}
K_{I\bar{J}\bar{K}}\Phi^{I}\bar{\Phi}^{\bar{J}}\bar{\Phi}^{\bar{K}}\cr
&\hskip -30pt+{1\over 3!}
K_{IJK\bar{L}}\Phi^{I}\Phi^{J}\Phi^{K}\bar{\Phi}^{\bar{L}}
+{1\over (2!)^2}
K_{IJ\bar{K}\bar{L}}\Phi^{I}\Phi^{J}\bar{\Phi}^{\bar{K}}\bar{\Phi}^{\bar{L}}
+{1\over 3!}
K_{I\bar{J}\bar{K}\bar{L}}\Phi^{I}\bar{\Phi}^{\bar{J}}
    \bar{\Phi}^{\bar{K}}\bar{\Phi}^{\bar{L}}\cr
&+\cdots}
}
where the coefficients $K_{I_1,I_2,\cdots,\bar{J}_1,\bar{J}_2,\cdots}$
are given by taking derivatives of the Kahler potential~:
\eqn\defKcoeffs{
K_{I_1,I_2,\cdots,\bar{J}_1,\bar{J}_2,\cdots}
{\partial\over\partial \Phi_{I_1}}
{\partial\over\partial \Phi_{I_2}}
\cdots
{\partial\over\partial \bar{\Phi}_{\bar{J}_1}}
{\partial\over\partial \bar{\Phi}_{\bar{J}_2}}
\cdots
K(\Phi,\bar{\Phi})\biggr|_{\Phi = \Phi_{cl}}.
}
We have dropped terms that involve only chiral or only antichiral
fields in \Kahlmet . The fact that the world sheet has collapsed to a
single space time point means that the coefficients of the power
series are constant and thus that in the action \Kahlact\ such terms
can be written as total derivatives. 

Note that it is not possible to use normal coordinates since the field
redefinitions necessary to transform to normal coordinates would in
general mix chiral and antichiral fields. The individual coefficients
are thus not covariant. As we will see below, however, the coefficients
nevertheless combine together to give a covariant result for the
Zamolodchikov metric.

Below we give the expressions for the Kahler metric, connection and
curvature tensor. 
The Kahler metric $g_{I\bar{J}}$ is given by 
\eqn\Kahlmet{
g_{I\bar{J}}= K_{I\bar{J}}.
}
Its inverse we denote by $K^{I\bar{J}}$.
The only non zero components of the connection are 
\eqn\Kahlconnect{
\Gamma^I_{JK}=K^{I\bar{L}}K_{\bar{L}JK}
\quad\quad{\rm and }\quad\quad
\Gamma^{\bar{I}}_{\bar{J}\bar{K}}=K^{\bar{I}L}K_{L\bar{J}\bar{K}}.
}
The curvature tensor $R_{I\bar{J}K\bar{L}}$ takes the simple form
\eqn\KahlR{
R_{I\bar{J}K\bar{L}}=K_{I\bar{J}K\bar{L}}
-K_{IK\bar{M}}K_{\bar{J}\bar{L}N}K^{\bar{M}N}.
}

\newsec{$N=2$ superfields and feynman rules in the prescence of a boundary}
In this section we fix our conventions for definitions of superfields
and superderivatives, and derive the superfield propagator and
superderivative/propagator identities in the presence of a boundary.

There are four fermionic variables
$\theta^+$, $\theta^-$, $\bar{\theta}^+$ and $\bar{\theta}^-$, and
a complex coordinate $z=x+iy$. Integration over $\theta$ and $z$ are given by
\eqn\thetazint{
\int d^4\theta = 
\int d\theta^+d\theta^-d\bar{\theta}^+d\bar{\theta}^-
\quad{\rm with}\quad
\int d\theta \,\theta = 1
\quad\quad{\rm and}\quad\quad
\int d^2z = \int d^2x
}
There are four superderivatives~:
\eqn\superD{\eqalign{
D^+&={\partial\over\partial\theta^+}+\bar{\theta}^+\bar{\partial}
\quad\quad
\bar{D}^+={\partial\over\partial\bar{\theta}^+}+\theta^+\bar{\partial}\cr
D^-&={\partial\over\partial\theta^-}+\bar{\theta}^-\partial
\quad\quad
\bar{D}^-={\partial\over\partial\bar{\theta}^-}+\theta^-\partial
}
}
They satisfy
\eqn\Dids{
\{D^+,\bar{D}^+\} = 2\bar{\partial},
\quad\quad
\{D^-,\bar{D}^-\} = 2\partial
\quad\quad{\rm and}\quad\quad
[D^2,\bar{D}^2] = -4\bar{\partial}\partial,
}
where we are using the conventions
\eqn\squares{
D^2=D^+D^-
\quad\quad{\rm and}\quad\quad
\theta^2=\theta^+\theta^-
}
and similarly for the barred superderivatives and $\theta$'s.

A chiral field $\Phi$ satisfies $\bar{D}\Phi(z,\theta) = 0$ and an 
antichiral field $\bar{\Phi}$ satisfies $D\bar{\Phi}(z,\theta) = 0$.
Their $\theta$ component expansions are
\eqn\chiralP{\eqalign{
\Phi(z,\theta)&=
\bigl[ 1
 +(\theta^-\bar{\theta}^-\partial+\theta^+\bar{\theta}^+\bar{\partial})
 -\theta^2\bar{\theta}^2\bar{\partial}\partial\bigr]X(z)\cr
&+
 \theta^+\bigl[1+\theta^-\bar{\theta}^-\partial\bigr]\Psi_+(z)
 +\theta^-\bigl[1+\theta^+\bar{\theta}^+\bar{\partial}\bigr]\Psi_-(z)\cr
&+
 \theta^2F(z)\cr
}
}
\eqn\antichiralP{\eqalign{
\bar{\Phi}(z,\theta)&=
\bigl[ 1
 -(\theta^-\bar{\theta}^-\partial+\theta^+\bar{\theta}^+\bar{\partial})
 -\theta^2\bar{\theta}^2\bar{\partial}\partial\bigr]X^*(z)\cr
&+
 \bar{\theta}^+
 \bigl[1-\theta^-\bar{\theta}^-\partial\bigr]\Psi_+^*(z)
 +\bar{\theta}^-
 \bigl[1-\theta^+\bar{\theta}^+\bar{\partial}\bigr]\Psi_-^*(z)\cr
&+
 \bar{\theta}^2F^*(z)\cr
}
}
We derive the propagator in flat space for a single chiral and
antichiral field. The propagator of the curved space action \Kahlact\
is then given by introducing indices $I$, $\bar{J}$ for the chiral and
antichiral fields and prefactoring the propagator we find below by the
inverse metric $K^{I\bar{J}}$. 

The flat space action for a single chiral and antichiral field is
\eqn\Squad{\eqalign{
S&=\int d^2z \,d^4\theta \,\Phi\bar{\Phi}\cr
 &=\int d^2z \,\bigl[X(-4\partial\bar{\partial})X^*
   +2\Psi_+\partial\Psi_+^*+2\Psi_-\bar{\partial}\Psi_-^*
   +|F|^2\bigr]\cr
 &=\int d^2x \,\bigl[X(-\nabla)X^*
   +\Psi_+(\partial_x-i\partial_y)\Psi_+^*
   +\Psi_-(\partial_x+i\partial_y)\Psi_-^*
   +|F|^2\bigr]\cr
}
}
The boundary conditions on the fields are
\eqn\bcs{
X(\bar{z},z) = - X(z,\bar{z})
\quad\quad{\rm and}\quad\quad
\Psi_-(z,\bar{z}) = - \Psi_+(\bar{z},z) 
}
This leads to the 
bosonic propagator
\eqn\bosprop{
<X(z_1,\bar{z}_1)X^*(z_2,\bar{z}_2)> 
= -{1\over 2\pi}\bigl(\ln|z_1-z_2|-\ln|z_1-\bar{z}_2|\bigr),
}
satisfying
\eqn\bospropid{\eqalign{
-4\partial_1\bar{\partial}_1<X(z_1,\bar{z}_1)X^*(z_2,\bar{z}_2)>
&= -\nabla<X(z_1,\bar{z}_1)X^*(z_2,\bar{z}_2)>\cr
&=\delta^2(z_1-z_2)-\delta^2(z_1-\bar{z}_2),
}
}
and fermionic propagators
\eqn\fermprop{\eqalign{
<\Psi_+(z)\Psi_+(z')>& = -{1\over 2\pi}{1\over \bar{z} - \bar{z}'}\cr
<\Psi_-(z)\Psi_-(z')>& = -{1\over 2\pi}{1\over z - z'}\cr
<\Psi_+(z)\Psi_-(z')>& = +{1\over 2\pi}{1\over \bar{z} - z'}\cr
<\Psi_-(z)\Psi_+(z')>& = +{1\over 2\pi}{1\over z - \bar{z}'}\cr
}
}

The superfield propagators can be built up from the component
propagators \bospropid\ and \fermprop . One finds (see the appendix for
some useful identities for the superderivatives).
\eqn\prop{\eqalign{
<\Phi(z_1,\theta_1)\bar{\Phi}(z_2,\theta_2)>
&=-{1\over 2\pi}D_1^2\bar{D}_2^2
[(\theta_1-\theta_2)^4\ln|z_1-z_2|
   +(\theta_1-\thetaswap_2)^4\ln|z_1-\bar{z}_2|]\cr
&=-{1\over 2\pi}D_1^2\bar{D}_1^2
[(\theta_1-\theta_2)^4\ln|z_1-z_2|
   -(\theta_1-\thetaswap_2)^4\ln|z_1-\bar{z}_2|]\cr
&=-{1\over 2\pi}\bar{D}_2^2D_2^2
[(\theta_1-\theta_2)^4\ln|z_1-z_2|
   -(\theta_1-\thetaswap_2)^4\ln|z_1-\bar{z}_2|]\cr
}
}
where $\tthetaswap$ means the $+$ and $-$ components have been 
interchanged. 

Again using the identities in the appendix one can show that the
propagators satisfy
\eqn\Dsqprop{\eqalign{
D_2^2<\Phi(z_1,\theta_1)\bar{\Phi}(z_2,\theta_2)>
=&D_2^2[(\theta_1-\theta_2)^4\delta^2(z_1-z_2)
   -(\theta_1-\thetaswap_2)^4\delta^2(z_1-\bar{z}_2)]\cr
\bar{D}_1^2<\Phi(z_1,\theta_1)\bar{\Phi}(z_2,\theta_2)>
=&\bar{D}_1^2[(\theta_1-\theta_2)^4\delta^2(z_1-z_2)
   -(\theta_1-\thetaswap_2)^4\delta^2(z_1-\bar{z}_2)]}
}
It is not so obvious that these identities would still hold true in the
presence of a boundary, (indeed for the $N=1$ superfield formalism the
analogous identities no longer hold when there is a boundary). The fact
that they do hold means that in evaluating Feynman diagrams one can
manipulate superderivatives and collapse propagators just as one does
for the case without boundary.

Finally we give the expressions for the propagators connected to the 
boundary operator 
${\cal O}(x) = \partial_y X(z)|_{y=0}$~:
\eqn\boundprop{\eqalign{
<{\cal O}(x_0)\bar{\Phi}(z_1,\theta_1)>
&=-{1\over\pi}\bar{D}_1^2\bar{\theta}_1^2 \partial_{y_1}\ln|x_0-z_1|,\cr
<{\cal O}^*(x_0)\Phi(z_1,\theta_1)>
&=-{1\over\pi}D_1^2\theta_1^2 \partial_{y_1}\ln|x_0-z_1|,
}
}
and for the tadpole propagator which starts and finishes at the same 
point
\eqn\tadprop{
<\Phi(z,\theta)\bar{\Phi}(z,\theta)>
=-{1\over 2\pi}\biggl[
\ln|0|-\ln|2y|
-i(\bar{\theta}^-\theta^+-\bar{\theta}^+\theta^-){1\over 2y}
+\theta^4\Bigr[{1\over 4y^2}-\pi\delta(0)\delta(2y)\Bigr]\biggr].
}
The tadpole propagator satisfies
\eqn\tadpropid{
D^2<\Phi(z,\theta)\bar{\Phi}(z,\theta)>^n
=\bar{D}^2<\Phi(z,\theta)\bar{\Phi}(z,\theta)>^n
=0,
}
where $n$ is any positive integer.
\subsec{Feynman rules}
The propagators given above were for a single chiral anti-chiral field
pair in flat space. The curved space propagator is given by including
chiral and anti-chiral indices and an inverse metric
$K^{I\bar{J}}$. Since the superderivatives act at opposite ends of
the propagator it is a standard convention to include the
superderivatives on the vertices rather than the propagators. In
other words diagrammatically we have for the propagators and vertices.
\eqn\propvert{\eqalign{
\epsfbox{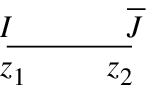}\,\,=&\,\,K^{I\bar{J}}{-1\over 2\pi}
[\delta(\theta_1-\theta_2)\ln|z_1-z_2|+
\delta(\theta_1-\thetaswap_2)\ln|z_1-\bar{z}_2|],\cr
\epsfbox{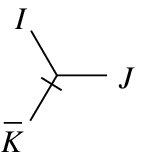}\,\,=&\,\,-K_{IJ\bar{K}}
(D^2\cdots)(D^2\cdots)(\bar{D}^2\cdots)
}
}
\vskip 15pt\hskip -19.5pt
The dots after the the $D^2$ and $\bar{D}^2$ for the vertices mean
that the superderivatives act on the propagators attached to the
vertices. For compactness of diagrammatic notation a solid bar on the
leg of a vertex denotes a $\bar{D}^2$ whereas for legs without a bar
we associate a $D^2$.

There are three diagrammatic rules that are easily derived
using the Feynman rules \propvert\ and the identities \Dsqprop .
Firstly each time one has a vertex with a single chiral field and all
other fields anti-chiral one can integrate by parts two
superderivatives off the antichiral legs onto the chiral leg. By the
identity \Dsqprop\ this collapses a propagator.
\eqn\identityI{
\hskip 50pt
\epsfbox{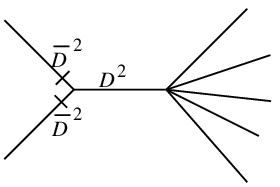}
\,\,\,=\,\,\,\epsfbox{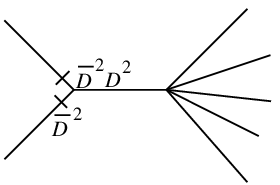}
\,\,\,=\,\,\,\epsfbox{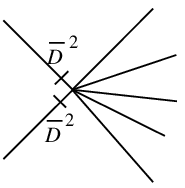}
}
\vskip 15pt\hskip -19.5pt
An analogous identity obviously applies when there is a single chiral
leg and all the other legs are chiral.

A second observation helps to reduce the number of diagrams that
contribute to the Zamolodchikov metric. For non-zero diagrams the
internal legs of the vertices connected to the boundary must have at
least one field of opposite chirality to the boundary
operator. Diagrams in which all legs are of the same chirality as
that of the boundary operator give zero since one can integrate by
parts superderivatives from the internal legs onto the external leg,
collapsing a propagator. The internal legs of the vertex then have
their endpoints on the boundary where the propagators are zero.
\footnote{$^{\diamond}$}{Note that there is a normal derivative from
the boundary operator acting on one of the propagators (rendering it
non-zero) but since there is more than one propagator with its
endpoint on the boundary the total result is zero.}
\eqn\identityII{
\hskip 40pt
\epsfbox{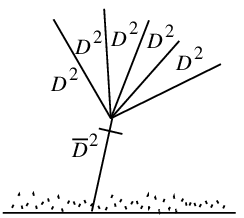}
\,\,\,=\,\,\,\epsfbox{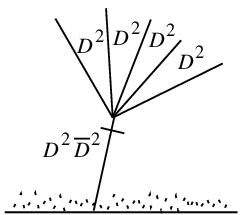}
\,\,\,=\,\,\,\epsfbox{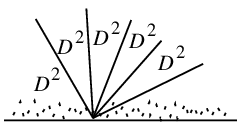}
\,\,\,=\,\,\,0
}
\vskip 10pt\hskip -19.5pt

Finally there is an identity involving the tadpole propagator which can 
be stated graphically as
\eqn\identityIII{
\raise -20pt\hbox{\epsfbox{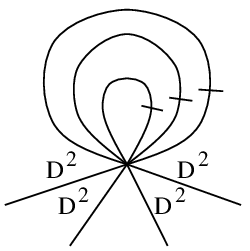}}\,\,=\,0.
}
The vertex to which the tadpoles are attached has all other legs of the
same chirality. One can thus integrate by parts a $D^2$ 
(or $\bar{D}^2$) off one of the legs and onto the tadpoles, leading to 
zero by \tadpropid .

\newsec{Zero, one and two loop contributions to the Zamolodchikov metric}
The tree level contribution to the Zamolodchikov metric is just given by the 
propagator. In other words we have 
\eqn\zeroloop{\eqalign{
g^{(0)}_{I\bar{J}}
&=2\pi(x_1-x_2)^2
\,<{\cal O}_I(x_1)\bar{{\cal O}}_{\bar{J}}(x_2)>_{\rm tree\,\,level}\cr
&=\,\,\hbox{\epsfbox{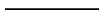}}\,\,K_{I\bar{J}}\cr
&=K_{I\bar{J}}
}
}
To simplify the presentation all Feynman diagrams in this section
include the prefactor $2\pi(x_1-x_2)^2$. The Feynman diagram
consisting of a single propagator 
\hbox{\epsfbox{zeroloop.eps}}
connecting the two boundary operators is thus, by definition, equal to
one.

Using the diagrammatic rule
\identityII\ their are only two diagrams that contribute at one loop~:
\eqn\oneloop{
g_{I\bar{I}}^{(1)}=
-K_{I\bar{I}J\bar{J}}K^{J\bar{J}}\epsfbox{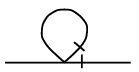}
+K_{IJ\bar{K}}K_{\bar{I}\bar{J}K}K^{J\bar{J}}K^{K\bar{K}}
\raise -10pt\hbox{\epsfbox{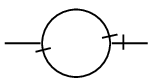}}
}
It is then easy to use the diagrammatic rule \identityI\ to collapse
the bottom propagator of the first diagram so that it has an identical
form to the first.
\eqn\oneloopresult{\eqalign{
g_{I\bar{I}}^{(1)}=&
\bigl[
-K_{I\bar{I}J\bar{J}}K^{J\bar{J}}
+K_{IJ\bar{K}}K_{\bar{I}\bar{J}K}K^{J\bar{J}}K^{K\bar{K}}
\bigr]\,\,
\raise -10pt\hbox{\epsfbox{oneloop1.eps}}\cr
=&-R_{I\bar{I}}\,\,
\raise -10pt\hbox{\epsfbox{oneloop1.eps}}\cr
=&\,\,0}
}
The tensors $K_{\cdots}$ have combined to give the Ricci tensor which is
zero at this order in $l_s^2$.

We now turn to the two loop diagrams. They involve vertices of order
three, four, five and six. Using the diagrammatic identity \identityI\
however all three vertices can have one of their propagators collapsed,
as can the four vertices with three legs of same chirality. One thus
finds that all diagrams collapse down to one of four distinct types.
We have
\eqn\twoloops{
g_{I\bar{I}}^{(2)}=g_{I\bar{I}}^{(2a)}+g_{I\bar{I}}^{(2b)}
+g_{I\bar{I}}^{(2c)}+g_{I\bar{I}}^{(2d)}.
}
The first type of contribution, $g_{I\bar{J}}^{(2a)}$ consists of
diagrams that collapse down to a double tadpole~:
\eqn\twoloopsa{
g_{I\bar{I}}^{(2a)}= g_{I\bar{I}}^{(2a1)}+g_{I\bar{I}}^{(2a2)}
+g_{I\bar{I}}^{(2a3)}+g_{I\bar{I}}^{(2a4)},
}
where
\eqn\twoloopsaone{
g_{I\bar{I}}^{(2a1)}
=-{1\over 2}K_{I\bar{I}K\bar{J}L\bar{K}}K^{J\bar{J}}K^{K\bar{K}}
\raise -10pt\hbox{\epsfbox{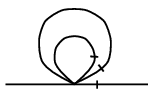}},
}
\eqn\twoloopsatwo{\eqalign{
g_{I\bar{I}}^{(2a2)}
=
&\biggr[
{1\over 2}K_{IJK\bar{L}}K_{\bar{I}\bar{J}\bar{K}L}
\raise -10pt\hbox{\epsfbox{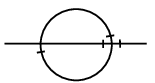}}
+K_{IJ\bar{K}L\bar{L}}K_{\bar{I}\bar{J}K}
\raise -10pt\hbox{\epsfbox{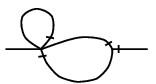}}\cr
&\hskip 10pt+K_{IJ\bar{K}}K_{\bar{I}\bar{J}KL\bar{L}}
\raise -10pt\hbox{\epsfbox{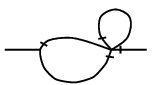}}
+{1\over 2}K_{I\bar{I}JK\bar{L}}K_{\bar{J}\bar{K}L}
\raise -10pt\hbox{\epsfbox{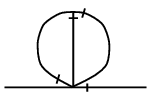}}\cr
&\hskip 10pt+{1\over 2}K_{I\bar{I}J\bar{K}\bar{L}}K_{\bar{J}KL}
\raise -10pt\hbox{\epsfbox{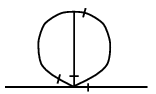}}
\biggr]
K^{J\bar{J}}K^{K\bar{K}}K^{L\bar{L}},
}}
\eqn\twoloopsathree{\eqalign{
g_{I\bar{I}}^{(2a3)}
=
&-\biggl[
K_{IJ\bar{K}}K_{\bar{I}L\bar{M}}K_{\bar{J}K\bar{L}M}
\raise -10pt\hbox{\epsfbox{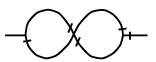}}
+K_{IJ\bar{K}}K_{\bar{I}\bar{J}L}K_{K\bar{L}M\bar{M}}
\raise -10pt\hbox{\epsfbox{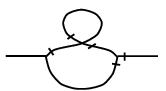}}\cr
&\hskip 10pt+{1\over 2}K_{I\bar{I}J\bar{K}}K_{\bar{J}L\bar{M}}K_{K\bar{L}M}
\raise -10pt\hbox{\epsfbox{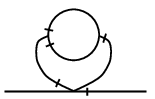}}
+K_{IJK\bar{L}}K_{\bar{I}\bar{J}M}K_{\bar{K}L\bar{M}}
\raise -10pt\hbox{\epsfbox{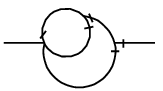}}\cr
&\hskip 10pt+{1\over 2}K_{IJK\bar{L}}K_{\bar{I}L\bar{M}}K_{\bar{J}\bar{K}M}
\raise -10pt\hbox{\epsfbox{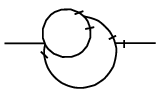}}
+K_{IJ\bar{K}\bar{L}}K_{\bar{I}L\bar{M}}K_{\bar{J}KM}
\raise -10pt\hbox{\epsfbox{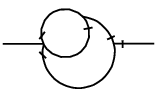}}\cr
&\hskip 10pt+K_{IJ\bar{K}}K_{\bar{I}K\bar{L}M}K_{\bar{J}L\bar{M}}
\raise -10pt\hbox{\epsfbox{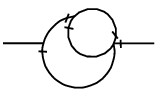}}
+{1\over 2}K_{IJ\bar{K}}K_{\bar{I}K\bar{L}\bar{M}}K_{\bar{J}LM}
\raise -10pt\hbox{\epsfbox{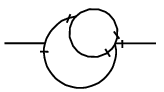}}\cr
&\hskip 10pt+K_{IJ\bar{K}}K_{\bar{I}\bar{J}\bar{L}M}K_{KL\bar{M}}
\raise -10pt\hbox{\epsfbox{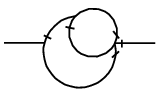}}
\biggr]
K^{J\bar{J}}K^{K\bar{K}}K^{L\bar{L}}K^{M\bar{M}},
}}
and 
\eqn\twoloopsafour{\eqalign{
g_{I\bar{I}}^{(2a4)}
=
&\biggl[
K_{IJ\bar{K}}K_{\bar{I}L\bar{M}}K_{\bar{J}\bar{L}N}K_{KM\bar{N}}
\raise -10pt\hbox{\epsfbox{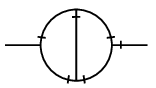}}
+K_{IJ\bar{K}}K_{\bar{I}L\bar{M}}K_{\bar{J}M\bar{N}}K_{K\bar{L}N}
\raise -10pt\hbox{\epsfbox{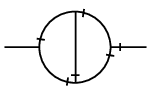}}\cr
&\hskip 10pt+K_{IJ\bar{K}}K_{\bar{I}L\bar{M}}K_{\bar{J}MN}K_{K\bar{L}\bar{N}}
\raise -10pt\hbox{\epsfbox{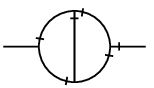}}
+K_{IJ\bar{K}}K_{\bar{I}\bar{J}L}K_{KM\bar{N}}K_{\bar{L}\bar{M}N}
\raise -10pt\hbox{\epsfbox{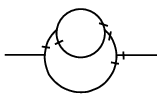}}\cr
&\hskip 10pt+{1\over 2}K_{IJ\bar{K}}K_{\bar{I}K\bar{L}}
K_{\bar{J}MN}K_{L\bar{M}\bar{N}}
\raise -10pt\hbox{\epsfbox{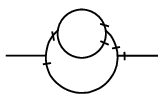}}
\biggr]
K^{J\bar{J}}K^{K\bar{K}}K^{L\bar{L}}K^{M\bar{M}}K^{N\bar{N}}.
}
}
The factors of $1/2$ are symmetry factors coming from symmetry under
interchange of two propagators. Using the diagrammatic rule \identityI\
all diagrams can be seen to reduce down to the double tadpole structure.
The non-covariant tensors $K_{\cdots}$ then combine to give the covariant 
result
\eqn\twoloopsaII{
g_{I\bar{I}}^{(2a)}
=-{1\over 2}\bigl[
\nabla_I\nabla_{\bar{I}}R
+R_{I\bar{J}K\bar{L}}R_{\bar{I}}^{\,\,\,\bar{J}K\bar{L}}
+R_{I\bar{I}J\bar{K}}R^{\bar{J}K}
\bigr]
\raise -10pt\hbox{\epsfbox{twoloops11.eps}}
}
The second type of contribution consists of all diagrams that collapse
down to a contraction between two four vertices, each with two chiral and
two antichiral indices
\eqn\twoloopsb{\eqalign{
g_{I\bar{J}}^{(2b)}
&=
{1\over 2}K_{IJ\bar{K}\bar{L}}K_{\bar{I}\bar{J}KL}
	K^{J\bar{J}}K^{K\bar{K}}K^{L\bar{L}}
\raise -10pt\hbox{\epsfbox{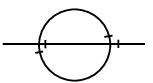}}\cr
&-\biggl[{1\over 2}K_{IJ\bar{K}\bar{L}}K_{\bar{I}\bar{J}M}K_{KL\bar{M}}
\raise -10pt\hbox{\epsfbox{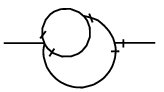}}\cr
&\hskip 10pt+{1\over 2}K_{IJ\bar{K}}K_{\bar{I}\bar{J}LM}K_{K\bar{L}\bar{M}}
\raise -10pt\hbox{\epsfbox{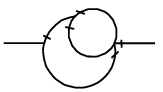}}
\biggr]
K^{J\bar{J}}K^{K\bar{K}}K^{L\bar{L}}K^{M\bar{M}}\cr
&+{1\over 2}K_{IJ\bar{K}}K_{\bar{I}\bar{J}L}K_{K\bar{M}\bar{N}}K_{\bar{L}MN}
	K^{J\bar{J}}K^{K\bar{K}}K^{L\bar{L}}K^{M\bar{M}}K^{N\bar{N}}
\raise -10pt\hbox{\epsfbox{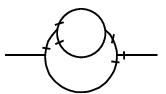}}\cr
&={1\over 2}R_{I\bar{J}K\bar{L}}R_{\bar{I}}^{\,\,\,\bar{J}K\bar{L}}
\raise -10pt\hbox{\epsfbox{twoloops21.eps}}
}
}
For the third type of contribution the diagrams collapse down to the
another possible contraction of two four vertices.
\eqn\twoloopsc{\eqalign{
g_{I\bar{J}}^{(2c)}
&=
K_{I\bar{I}J\bar{K}}K_{\bar{J}KL\bar{L}}
	K^{J\bar{J}}K^{K\bar{K}}K^{L\bar{L}}
\raise -10pt\hbox{\epsfbox{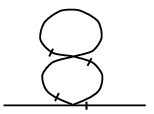}}\cr
&-\biggl[
K_{IJ\bar{K}}K_{\bar{I}K\bar{L}}K_{\bar{J}LM\bar{M}}
\raise -10pt\hbox{\epsfbox{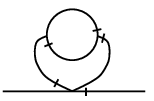}}\cr
&\hskip 10pt+K_{I\bar{I}J\bar{K}}K_{\bar{J}L\bar{M}}K_{K\bar{L}M}
\raise -10pt\hbox{\epsfbox{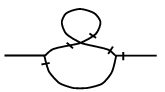}}
\biggr]
K^{J\bar{J}}K^{K\bar{K}}K^{L\bar{L}}K^{M\bar{M}}\cr
&+K_{IJ\bar{K}}K_{\bar{I}K\bar{L}}K_{\bar{J}M\bar{N}}K_{L\bar{M}N}
	K^{J\bar{J}}K^{K\bar{K}}K^{L\bar{L}}K^{M\bar{M}}K^{N\bar{N}}
\raise -10pt\hbox{\epsfbox{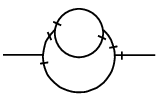}}\cr
&=R_{I\bar{I}J\bar{K}}R^{J\bar{K}}\raise -10pt\hbox{\epsfbox{twoloops31.eps}}
}
}
Finally there is the contribution consisting of two one loop diagrams
\eqn\twoloopsd{\eqalign{
g_{I\bar{J}}^{(2d)}
&=
K_{I\bar{J}K\bar{K}}K_{\bar{I}JL\bar{L}}
	K^{J\bar{J}}K^{K\bar{K}}K^{L\bar{L}}
\raise -10pt\hbox{\epsfbox{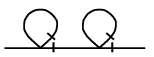}}\cr
&-\biggl[K_{I\bar{J}K\bar{K}}K_{\bar{I}L\bar{M}}K_{J\bar{L}M}
\raise -10pt\hbox{\epsfbox{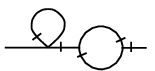}}\cr
&\hskip 10pt+K_{IJ\bar{K}}K_{\bar{I}LM\bar{M}}K_{\bar{J}K\bar{L}}
\raise -10pt\hbox{\epsfbox{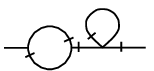}}
\biggr]	K^{J\bar{J}}K^{K\bar{K}}K^{L\bar{L}}K^{M\bar{M}}\cr
&+K_{IJ\bar{K}}K_{\bar{I}L\bar{M}}K_{\bar{J}K\bar{N}}K_{\bar{J}K\bar{N}}
	K^{J\bar{J}}K^{K\bar{K}}K^{L\bar{L}}K^{M\bar{M}}K^{N\bar{N}}
\raise -10pt\hbox{\epsfbox{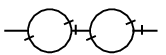}}\cr
&=R_{I\bar{J}}R_{\bar{I}}^{\,\,\,J}\raise -10pt\hbox{\epsfbox{twoloops41.eps}}
}
}
If one was confident that the result would be covariant one
could specialise to Kahler potentials where all three vertices are
zero, calculate using only the four vertices and the six vertex
and find directly the covariant results of 
\twoloopsaII\twoloopsb\twoloopsc\twoloopsd\
It can be shown \grisaru\ that the divergent part will always be
covariant. The proof of \grisaru\ does not however apply to the finite
part of the boundary two point function.

For Ricci flat metrics the first and third terms in \twoloopsa\ are
zero as are the contributions \twoloopsc\ and \twoloopsd .
For Ricci flat metrics we thus have
\eqn\twoloopsabcd{
g_{I\bar{J}}^{(2)}
={1\over 2}R_{I\bar{J}K\bar{L}}R_{\bar{I}}^{\,\,\,\bar{J}K\bar{L}}
\biggl[
-\raise -10pt\hbox{\epsfbox{twoloops11.eps}}
+\raise -10pt\hbox{\epsfbox{twoloops21.eps}}
\biggr]
}
Note that the second term cannot be further reduced using
\identityI . One can still integrate by parts the superderivatives
to collapse an internal propagator and end up with double tadpole
structure of the first term (thus canceling the divergence). In so
doing however one will also end up with (in addition to the double
tadpole structure) terms involving superderivatives acting on the
boundary propagators. It is these terms (which are finite) which give
the contribution to the Zamolodchikov metric. The whole contribution
to the Zamolodchikov metric comes from the finite part of the second
term. We thus see that the Zamolodchikov metric potentially receives a
contribution proportional to
$R_{I\bar{J}K\bar{L}}R_{\bar{I}}^{\,\,\,\bar{J}K\bar{L}}$.  All that
remains to do is to calculate the precise coefficient that goes with
this term.
\subsec{Calculation of coefficient of $R^2$ term}
Below we indicate explicitly the superderivatives for the double four
vertex term.
\eqn\doublefour{
\raise -20pt\hbox{\epsfbox{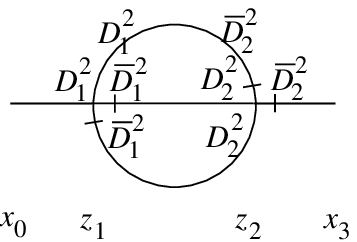}}
}

To reproduce the tadpole structure one can integrate by parts the
$\bar{D}_1^2$ superderivatives off the bottom propagator. This will
generate several types of contribution. There will be terms with
superderivatives acting on the left hand external propagator. There
will also be a contribution in which both $\bar{D}_1^2$ act on the
top propagator. This is equivalent (via identity \Dsqprop ) to
$\bar{D}_1^2$ acting on a collapsed propagator. Integrating the
superderivatives back off the collapsed propagator leaves one with the
double tadpole structure and in addition further terms in which
superderivatives act on the external propagator.\footnote{$^{\star}$}
{Note that if one was calculating the beta function all terms
with superderivatives acting on the external legs would be dropped
since, by power counting arguments, they give rise to finite
contributions.} Alternatively and more simply one can manipulate
directly the expression for the top propagator and express it as a
collapsed propagator + other terms. The identity
\eqn\superidentity{
D_1^2\bar{D}_1^2=\bar{D}_1^2D_1^2 -4\partial_1\bar{\partial}_1
+2(\bar{\partial}_1\bar{D}_1^-D_1^-+\partial_1\bar{D}_1^+D_1^+),
}
along with the expressions \prop\ for the
propagators  means that we can rewrite the diagram as follows
\eqn\fourfourvertex{
\raise -20pt\hbox{\epsfbox{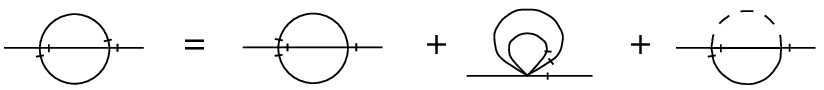}}
}
where the three diagrams on the right hand side of \fourfourvertex\
correspond, respectively to the three terms on the right hand side of
\superidentity . In particular the dotted line of the top propagator
of the third diagram comes from the third term of \superidentity .

The first diagram of\fourfourvertex\ is zero by \identityII , the
second leads to cancellation of the double tadpole in
\twoloopsabcd\ leaving the final diagram. Writing out the final
diagram explicitly we have
\eqn\finitetwoloopsupper{\eqalign{
\raise -7pt\hbox{\epsfbox{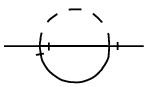}}=
4\pi|x_0-x_3|^2
\int d^2z_1 d^4\theta_1
\int d^2z_2 d^4\theta_2
&\bigl[(D_1^2P_{10}) (\bar{D}_2^2P_{23}) (\bar{D}_1^2D_1^2P_{12})^2\cr
&\hskip 40pt
(\bar{\partial}_1\bar{D}_1^-D_1^-+\partial_1\bar{D}_1^+D_1^+)P_{12}
\bigr],}
}
where 
\eqn\defPs{\eqalign{
P_{12}&=-{1\over 2\pi}
\bigl[
\delta^4(\theta_1-\theta_2)\ln|z_1-z_2|
-\delta^4(\theta_1-\thetaswap_2)\ln|z_1-\bar{z}_2|
\bigr]\cr
P_{10}&=-{1\over\pi}\theta_1^2\partial_{y_1}\ln|z_1-x_0|\cr
P_{23}&=-{1\over\pi}\bar{\theta}_2^2\partial_{y_1}\ln|z_2-x_3|}
}
As discussed for the zero loop contribution we include in the definition
of the Feynman diagrams a prefactor $2\pi|x_0-x_3|^2$.
The integrals over $z_1$ and $z_2$ in \finitetwoloopsupper\ 
are over the upper half complex
plane. By interchanging $z_1$ with $\bar{z}_1$, $\theta_1^+$ with
$\theta_1^-$ and  $\bar{\theta}_1^+$ with $\bar{\theta}_1^-$ the
integral over $z_1$ can be completed into an integral over the whole
complex $z_1$ plane. After using the identities
\eqn\idenityIII{\eqalign{
\bar{\partial}_1D_1^-\delta^4(\theta_1-\theta_2)f(z_1-z_2)=&
\bar{\partial}_2D_2^-\delta^4(\theta_1-\theta_2)f(z_1-z_2)\cr
\bar{\partial}_1D_1^-\delta^4(\theta_1-\thetaswap_2)f(z_1-\bar{z}_2)=&
\partial_2D_2^+\delta^4(\theta_1-\thetaswap_2)f(z_1-\bar{z}_2)},
}
where $f(z)$ is an arbitrary function of $z$ and its complex conjugate,
the $z_2$ integral can similarly be completed into an integral over
the whole complex $z_2$ plane. We thus arrive at the integral
\eqn\dtwozdfourthsq{\eqalign{
\raise -7pt\hbox{\epsfbox{twoloops52.eps}}=&
|x_0-x_3|^2
\int d^2z_1  d^4\theta_1 \int d^2z_2 d^4\theta_2
\biggl[(\bar{D}_1^-D_1^2 P_{10})(D_2^-\bar{D}_2^2 P_{23})\cr
&\hskip 160pt(\bar{D}_1^2D_1^2 P_{12})^2
\delta(\theta_1-\theta_2)^4{1\over \bar{z}_1 - \bar{z}_2}\biggr], }
}
where the integrals are now over the whole complex plane.
Performing the integrations over the fermionic parameters we find 
(see appendix)
\eqn\doublecomplane{
\raise -7pt\hbox{\epsfbox{twoloops52.eps}}=
{1\over 2\pi^4}\int d^2z_1 d^2z_2
\biggl({1\over z_1^2}-{1\over \bar{z}_1^2}\biggr){1\over z_2^2}
{1\over \bar{z}_1-\bar{z}_2+1}{1\over \bar{z}_1-z_2+1}
\ln|z_1-z_2+1|
}
Evaluating the double integral over the complex plane one finds
(see appendix)
\eqn\Rsqcoeff{
\raise -7pt\hbox{\epsfbox{twoloops52.eps}}={1\over 6}.
}
Up to two loop order the D0 metric is thus given by~:
\eqn\Dzerometric{
g^D_{I\bar{I}}=g^{\sigma}_{I\bar{I}}+{l_s^4\over 12}
R_{I\bar{J}K\bar{L}}R_{\bar{I}}^{\,\,\,\bar{J}K\bar{L}}
+{\cal O}(l_s^6).
}

\newsec{Conclusions}
The conclusion of this paper is that there is a non-trivial
contribution to the Zamolodchikov metric at order $l_s^4$. A D-brane
in a weakly curved background thus experiences a different metric
from that seen by the string.

\newsec{Acknowledgements} I would like to thank Mike Douglas for
pointing out the interest of this problem and for discussion in the
early stages of this work. I also especially thank Philippe Brax for
collaboration on an early version of this project.  

\newsec{Appendix}
In this appendix we list the identities necessary to prove
the results of section 3. We also give a few technical details on the 
calculation of the coefficient of the $R^2$ term.

\subsec{Useful identities}

\eqn\Dexpid{\eqalign{
&D_1^2(\theta_1-\theta_2)^2=-\exp
[\bar{\theta}_1^-(\theta_1^--\theta_2^-)\partial
+\bar{\theta}_1^+(\theta_1^+-\theta_2^+)\bar{\partial}]\cr
&\bar{D}_1^2(\bar{\theta}_1-\bar{\theta}_2)^2=-\exp
[\theta_1^-(\bar{\theta}_1^--\bar{\theta}_2^-)\partial
+\theta_1^+(\bar{\theta}_1^+-\bar{\theta}_2^+)\bar{\partial}]\cr
&\bar{D}_2^2D_1^2(\theta_1-\theta_2)^4 =\exp
[\bar{\theta}_1^-\theta_1^-\partial_1
   +\bar{\theta}_1^+\theta_1^+\bar{\partial}_1
-\bar{\theta}_2^-\theta_2^-\partial_2
   -\bar{\theta}_2^+\theta_2^+\bar{\partial}_2\cr
&\hskip 140pt-\bar{\theta}_1^-\theta_2^-(\partial_1-\partial_2)
   -\bar{\theta}_1^+\theta_2^+(\bar{\partial}_1-\bar{\partial}_2)]\cr
&\bar{D}_2^2D_1^2(\theta_1-\thetaswap_2)^4 =-\exp
[\bar{\theta}_1^-\theta_1^-\partial_1
   +\bar{\theta}_1^+\theta_1^+\bar{\partial}_1
-\bar{\theta}_2^-\theta_2^-\partial_2
   -\bar{\theta}_2^+\theta_2^+\bar{\partial}_2\cr
&\hskip 140pt-\bar{\theta}_1^-\theta_2^+(\partial_1-\bar{\partial}_2)
   -\bar{\theta}_1^+\theta_2^-(\bar{\partial}_1-\partial_2)]\cr
&D_1^2(\theta_1-\theta_2)^4 f(z_1-z_2) = 
D_2^2(\theta_1-\theta_2)^4 f(z_1-z_2)\cr
&D_1^2(\theta_1-\thetaswap_2)^4 f(z_1-\bar{z}_2) = 
-D_2^2(\theta_1-\thetaswap_2)^4 f(z_1-\bar{z}_2)\cr
}
}
where $f(z)$ is an arbitrary function of $z$ and $\bar{z}$ and 
$\tthetaswap$ means the $+$ and $-$ components have been 
interchanged.
\subsec{Fermionic integrals for $R^2$ contribution}
To evaluate the fermionic integrals of equation\dtwozdfourthsq\ one
first trivially integrates over $\theta_2$, the delta function setting
all occurrences of $\theta_2$ equal to $\theta_1$. One then
uses the following expressions for the propagators (which follow from
\defPs\ and \Dexpid )~:
\eqn\Pids{\eqalign{
\bar{D}_1^-D_1^2 P_{10}=&\,\,\,\,
{1\over\pi}\partial_{y_1}D_1^+\theta_1^2{1\over z_1-x_0},\cr
D_2^-\bar{D}_2^2 P_{23}=&\,\,\,\,
{1\over\pi}\partial_{y_2}\bar{D}_2^+\bar{\theta}_2^2{1\over z_2-x_3},\cr
\bar{D}_1^2D_1^2 P_{12}|_{\theta_1=\theta_2}=&-{1\over 2\pi}\biggr[
\ln|z_1-z_2|-\ln|z_1-\bar{z}_2|\cr
&\hskip 20pt
+\bar{\theta}_1^-\theta_1^+{1\over z_1-\bar{z}_2}
+\bar{\theta}_1^+\theta_1^-{1\over \bar{z}_1-z_2}\cr
&\hskip 20pt
-(\bar{\theta}_1^+\theta_1^++\bar{\theta}_1^-\theta_1^-){1\over 2}
\Bigr[{1\over z_1-\bar{z}_2}+{1\over \bar{z}_1-z_2}\Bigr]\cr
&\hskip 20pt -\theta_1^4\Bigr[{1\over (z_1-\bar{z_2})^2}+{1\over
(\bar{z}_1-z_2)^2} +\pi\delta^2(z_1-\bar{z}_2)\Bigl]\biggl].  
}  
}
Integrating over the fermionic parameters $\theta_1$ there are two
potential contributions. In the first all four powers of $\theta$ come
from the external propagators $(\bar{D}_1^-D_1^2 P_{10})$ and
$(D_2^-\bar{D}_2^2 P_{23})$ and none from the $(\bar{D}_1^2D_1^2
P_{12})^2$ term. Such contributions arise from the spatial derivative
part of the $\bar{D}_1$ or $D_2^-$ superderivative. In either case
one of the external propagators is collapsed and either the $z_1$ or $z_2$
vertex is pulled back to the boundary, where the $(\bar{D}_1^2D_1^2
P_{12})^2$ term gives zero.  For the second type of contribution just
two $\theta$'s come from the external propagators
($\theta^-\bar{\theta}^-$), the remaining two coming from the
$(\bar{D}_1^2D_1^2 P_{12})^2$ term. This leads to \doublecomplane .

\subsec{Double integration over the complex plane}
The double integral over the complex plane \doublecomplane\ is
complicated by the fact that it mixes holomorphic and antiholomorphic
variables. Below we describe briefly how the integral can be performed
analytically and give an intermediate result which allows the
integral to be checked by numerical integration. 

Changing integration variables from $z_1$, $z_2$ to 
$z^{\pm}=(z_1\pm z_2)/2$ the integral \doublecomplane\ can be written
in the form
\eqn\zplusminusint{
\raise -7pt\hbox{\epsfbox{twoloops52.eps}}=
{2\over \pi^4}\int d^2z^-\Bigl[I_1(z^-)-I_2(z^-)\Bigr]
{1\over 2\bar{z}^- + 1}\ln|2z^- + 1|,
}
with 
\eqn\Ionetwoint{\eqalign{
I_1(z^-)=&\int d^2z^+ {1\over (z^+ + z^-)^2}{1\over (z^+ - z^-)^2}
{1\over z^+-\bar{z}^++z^-+\bar{z}^-+1},\cr
I_2(z^-)=&\int d^2z^+ {1\over (\bar{z}^+ + \bar{z}^-)^2}{1\over (z^+ - z^-)^2}
{1\over z^+-\bar{z}^++z^-+\bar{z}^-+1}.\cr
}
}
Performing the integrals over $z^+$ one finds the results
\eqn\Ionetwo{\eqalign{
I_1(z^-)=&-{i\pi\over 2{z^-}^3}\tan^{-1}\Bigl({y^-\over x^-+1/2}\Bigr),\cr
I_2(z^-)=&{\pi\over 4}\biggl[
{2\over (2x^-+1/2)^3}\bigl[\ln|2z^-|-\ln|2z^-+1|\bigr]\cr
&
-{1\over (2x^-+1/2)^2}\Bigl[{1\over z^-}+{1\over \bar{z}^-}\Bigr]\cr
&
-{1\over 2(2x^-+1/2)}\Bigl[{1\over (z^-)^2}+{1\over
(\bar{z}^-)^2}\Bigr]
\biggr].
}
}
The inverse tangent in $I_1(z^-)$ takes on values between $-\pi/2$ and
$\pi/2$. 
Note that the expression for $I_2(z^-)$ is non-singular
at $x^- = -1/4$ (contrary to first appearances). The 
singularities of the individual terms cancel among themselves leaving 
$I_2(z^-)$ finite at $x^- = -1/4$.

Integration over $z^-$ yields~:
\eqn\intIonetwo{\eqalign{
\int d^2z^-I_1(z^-){1\over 2\bar{z}^- + 1}\ln|2z^- + 1|,
=&{\pi^4\over 8}\cr
\int d^2z^-I_2(z^-){1\over 2\bar{z}^- + 1}\ln|2z^- + 1|
=&{\pi^4\over 24},
}
}
Which on substituting back into \zplusminusint\ leads to equation\Rsqcoeff .
\listrefs

\bye